\documentclass[10pt,conference]{IEEEtran}
\IEEEoverridecommandlockouts

\usepackage{cite}
\usepackage{amsmath,amssymb,amsfonts}
\usepackage{algorithmic}
\usepackage{graphicx}
\usepackage{textcomp}
\usepackage{xcolor}
\usepackage{listings}
\usepackage{verbatim}
\usepackage{pifont}
\usepackage{fancyvrb}
\usepackage{fancyhdr}
\usepackage{colortbl}
\usepackage{hhline}
\usepackage{tabularx}
\usepackage{subfig}
\usepackage{url}

\def\BibTeX{{\rm B\kern-.05em{\sc i\kern-.025em b}\kern-.08em
    T\kern-.1667em\lower.7ex\hbox{E}\kern-.125emX}}
\begin{document}

\title{An Analysis of the Automatic Bug Fixing Performance of ChatGPT}  

\author{
\IEEEauthorblockN{Dominik Sobania}
\IEEEauthorblockA{Johannes Gutenberg University Mainz\\
Email: dsobania@uni-mainz.de}\\  
\IEEEauthorblockN{Carol Hanna}
\IEEEauthorblockA{University College London\\
Email: carol.hanna.21@ucl.ac.uk}
\and
\IEEEauthorblockN{Martin Briesch}
\IEEEauthorblockA{Johannes Gutenberg University Mainz\\
Email: briesch@uni-mainz.de}\\
\IEEEauthorblockN{Justyna Petke}
\IEEEauthorblockA{University College London\\
Email: j.petke@ucl.ac.uk}
}

\maketitle

\begin{abstract}
To support software developers in finding and fixing software bugs, several automated program repair techniques have been introduced. Given a test suite, standard methods usually either synthesize a repair, or navigate a search space of software edits to find test-suite passing variants. Recent program repair methods are based on deep learning approaches. One of these novel methods, which is not primarily intended for automated program repair, but is still suitable for it, is ChatGPT. The bug fixing performance of ChatGPT, however, is so far unclear. Therefore, in this paper we evaluate ChatGPT on the standard bug fixing benchmark set, QuixBugs, and compare the performance with the results of several other approaches reported in the literature. We find that ChatGPT's bug fixing performance is competitive to the common deep learning approaches CoCoNut and Codex and notably better than the results reported for the standard program repair approaches. In contrast to previous approaches, ChatGPT offers a dialogue system through which further information, e.g., the expected output for a certain input or an observed error message, can be entered. By providing such hints to ChatGPT, its success rate can be further increased, fixing 31 out of 40 bugs, outperforming state-of-the-art.
\end{abstract}


\section{Introduction}

Complex software usually contains undiscovered bugs in its source code. 
The later these are found, the more far-reaching consequences these can have. 
Uncorrected bugs in software can lead to failures of essential systems, which can result in high economic costs \cite{wong2017more}. 

In order to support programmers in finding and fixing software errors, automated program repair (APR) systems have been introduced that automatically suggest software patches to correct the detected errors \cite{le2011genprog, gazzola2018automatic}. 
For instance, Haraldsson~et~al.~\cite{haraldsson2017fixing} suggest an approach based on genetic improvement (GI)~\cite{petke2017genetic} that tracks emerging bugs during a workday and searches for potential fixes for them overnight. 
The following morning the programmers get a list of suggestions which should help fix the detected bugs. 

Standard methods for automated program repair can be classified into two categories: the generate-and-validate approaches mutate software guided by a search strategy, while semantics-driven (or synthesis-based) approaches use a constraint solver to synthesize repairs~\cite{gazzola2018automatic}.
The generate-and-validate ones have first seen industrial uptake~\cite{haraldsson2017fixing}.
One of the key disadvantage of standard approaches to APR is their running cost.
The generate-and-validate ones usually rely on test suites to verify program correctness, while synthesis-based ones on calls to a constraint solver.
Both validation strategies are costly, making typical APR tools hours to run before a viable patch is presented to the developer.

Most recently, program repair tools based on deep learning~(DL) approaches have been introduced~\cite{zhang2023asurvey}. 
These learn bug fixing patterns from existing databases and treat the automated program repair problem as a neural machine translation task, producing a ranking of, sometimes hundreds of, patches.
Unlike standard approaches, such generated patches are not usually evaluated against a test suite, or other automated verification strategy, so may not even compile. 
Nevertheless, DL-based program repair has shown competitive results to standard approaches~\cite{zhang2023asurvey}.

In recent years, several large-scale language models based on the Transformer architecture \cite{vaswani2017attention} have been introduced, such as CodeBERT \cite{feng2020codebert}, PyMT5 \cite{clement2020pymt5}, and Codex \cite{chen2021evaluating}, which can also process and extend source code and achieve comparable results to standard approaches on various coding tasks \cite{sobania2022choose}. 
A large-scale language model based on the Transformer architecture that has recently received great attention is ChatGPT.\footnote{\url{https://openai.com/blog/chatgpt/} (accessed January 18, 2023).} 
With ChatGPT not only text input can be extended, but it is even possible to have a conversation with the language model and the previous chat history is taken into account for answer generation. 
In addition to very general or subject-specific topics, ChatGPT can also be used to discuss source code, e.g., to ask for a suggestion for a fix of incorrect code. 
However, the quality of these suggestions is still unclear.

Therefore, in this work we evaluate and analyse the automatic bug fixing performance of ChatGPT.
Moreover, we provide a comparison with results reported in the literature obtained using state-of-the-art APR approaches and Codex.
We chose the QuixBugs~\cite{lin2017quixbugs} benchmark set for our study, as it contains small, yet challenging programs for current APR approaches.
We consider all Python problems from QuixBugs, i.e., 40 overall.

We first ask ChatGPT for bug fixes for the selected benchmarks and manually check whether the suggested solution is correct or not.
We repeat the query four times, to account for the heuristic nature of ChatGPT.
Next, we compare its performance with that of Codex and dedicated APR approaches. 
For the standard APR approaches, we take the results from a recent paper \cite{ye2021comprehensive} that examines the performance of several methods on the QuixBugs benchmark set.
For dedicated APR methods based on deep learning, we take results from CoCoNut~\cite{lutellier2020coconut}.\footnote{Although more recent approaches exist, we found this work is the most recent providing sufficient patch ranking detail.}
For the large-scale language model Codex, we take the results from \cite{prenner2022can}.   
Furthermore, we study and categorize ChatGPT's answers to gain a deeper understanding of its behavior. 
Given that ChatGPT provides a unique opportunity for a conversation with the model, we provide a small hint to the model (e.g., a failing test input with an error it produces) to see if it improves ChatGPT's fix rate.

We find that ChatGPT's program repair performance is competitive to the results achieved with CoCoNut and Codex ($19$ vs. $19$ vs. $21$ instances solved, respectively). 
Compared to the standard program repair approaches, ChatGPT achieves notably better results. 
With ChatGPT, we could fix bugs in $19$ out of $40$ problems while with the standard approaches only $7$ can be fixed, even though we give ChatGPT only the incorrect code snippet without any additional information and without using the chat option in a conversational way. 
If the chat function is actively used, we can fix even more instances.
This shows the power of providing manual hints to a program repair system.
All our experimental data is available online.\footnote{\url{https://gitlab.rlp.net/dsobania/chatgpt-apr}.}

\section{ChatGPT for Automated Program Repair}\label{sec:method}

In this section we present our methodology for assessing ChatGPT's program repair performance.

\subsection{Benchamrk}

To evaluate the automatic bug fixing performance of ChatGPT, we use the QuixBugs \cite{lin2017quixbugs} benchmark set. 
Unlike many other benchmark suites for automated program repair, QuixBugs contains relatively small problems (small number of code lines).
These are thus suitable for use in a dialogue system. 
For each of the $40$ benchmark problems from QuixBugs, we take the erroneous Python code, remove all contained comments\footnote{This was necessary, as sometimes the comments contain the solution.}, and ask ChatGPT if the code contains a bug and how it can be fixed. 
For each benchmark problem, we make several independent requests to ChatGPT and manually check whether the given answer is correct or not. 
We standardize our procedure by using the same format for each query. We ask: ``Does this program have a bug? How to fix it?'' followed by an empty line and the buggy code without comments.
Figure~\ref{fig:example_request} shows an example request to ChatGPT for the {\sc bitcount} problem. 
\begin{figure}[!ht]
\begin{lstlisting}[frame=lines,numbers=left,xleftmargin=2.4em,framexleftmargin=2.3em,basicstyle=\small]
Does this program have a bug? How to 
fix it?

def bitcount(n):
    count = 0
    while n:
        n ^= n - 1
        count += 1
    return count
\end{lstlisting}
\caption{Request to ChatGPT for the {\sc bitcount} problem.}
\label{fig:example_request}
\end{figure}
Lines $1$-$2$ contain the question to ChatGPT where we ask how the bug can be fixed and starting from line $4$ we present the erroneous code snippet. 
For this example, we would expect from ChatGPT an answer that addresses the bug in line $7$, where \texttt{\lstinline{n ^= n - 1}} should be replaced with \texttt{\lstinline{n &= n - 1}}, either with a response containing the complete code snippet with the fixed bug (correctly addressed) or by giving an exact and correct description how to change the affected code lines.

\subsection{Comparison Study}

We ran four independent requests to ChatGPT for each problem from the QuixBugs dataset.
In order to compare the results of ChatGPT with the standard APR methods, we take the results from a comprehensive study from the literature \cite{ye2021comprehensive} that reports the performance of ten different methods (Arja~\cite{yuan2018arja}, Cardumen~\cite{martinez2018ultra}, Dynamoth~\cite{durieux2016dynamoth}, JGenProg~\cite{martinez2019astor}, JKali~\cite{martinez2019astor}, JMutRepair~\cite{martinez2019astor}, Nopol~\cite{xuan2016nopol}, NPEfix~\cite{cornu2015npefix}, RSRepair~\cite{yuan2018arja}, and Tibra~\cite{martinez2019astor}) on the problems from QuixBugs. 
For dedicated APR approaches based on deep learning we chose recent results reported by Lutellier~et~al.~\cite{lutellier2020coconut}.\footnote{CoCoNut, solves overall only 2 instances less than best reported thus far on the QuixBugs Python dataset\cite{prenner2022can}, though details on patch ranking for each program were missing from the later work.}
In Table~\ref{tab:results} we report a fix only if the correct patch was ranked first by Lutellier~et~al.'s proposed approach, CoCoNut.
For the large-scale language model Codex, we take the results from a recent paper \cite{prenner2022can}. 
We ran this experiment on ChatGPT versions from December 15, 2022 and January 9, 2023. 

\subsection{Dialogue Study}
\label{subsec:dialogue}

Given that ChatGPT provides a unique opportunity of a dialogue with the model, we also conduct a study where we provide ChatGPT with a hint, based on ChatGPT's response. 
If ChatGPT does not provide a correct answer to the first request (described in the previous paragraph), we tell ChatGPT in a standardized way that the function is not working correctly and additionally provide an input example that shows that the function is not working properly. 
If ChatGPT incorrectly claimed the program was correct, we replied:
``The function does not work. E.g., for the input $<$input$>$ it should return $<$output$>$.''
or ``The function does not work. E.g. for the input $<$input$>$  I get the following error message: $<$output$>$'', depending on whether the failing test case from the QuixBugs dataset returned an incorrect answer or threw an error.
In the case of more complex inputs we made the following response: ``The function does not work. E.g., given the following call: $<$code snippet$>$ The following should be the output: $<$output$>$.''\footnote{The third case only appeared once. All queries are available online.}
We only provide one such hint and report results.
This experiment was run on the ChatGPT version from January 9, 2023.

\definecolor{lightgray}{gray}{0.9}
\begin{table*}
  \centering
  \caption{Results achieved by ChatGPT, Codex, CoCoNut, and the standard APR approaches on the problems from the QuixBugs benchmark set. For ChatGPT, we also report the number of successful runs in brackets.}
  \renewcommand{\arraystretch}{1.64}
  \label{tab:results}
  \begin{tabular}{lcccc}
    \hline
    \textbf{Benchmark problem}\;\;\;\;\;\;\;\;\;\;\;\;\;\;\;\;\;\; & \;\;\;\;\textbf{ChatGPT}\;\;\;\; & \;\;\;\;\textbf{Codex \cite{prenner2022can}}\;\;\;\; & \;\;\;\;\textbf{CoCoNut \cite{lutellier2020coconut}}\;\;\;\; & \;\;\;\textbf{Standard APR \cite{ye2021comprehensive}}\;\;\; \\
    \hline
    bitcount                   & \ding{55} (0 / 4) & \ding{51} & \ding{51}  & \ding{55} \\
    \rowcolor{lightgray}
    breadth-first-search       & \ding{51} (2 / 4) & \ding{55} & \ding{51} & \ding{55} \\
    bucketsort                 & \ding{51} (4 / 4) & \ding{51} & \ding{51} & \ding{55} \\
    \rowcolor{lightgray}
    depth-first-search         & \ding{55} (0 / 4) & \ding{51} & \ding{55} & \ding{55} \\
    detect-cycle               & \ding{55} (0 / 4) & \ding{55} & \ding{55} & \ding{51} \\
    \rowcolor{lightgray}
    find-first-in-sorted       & \ding{51} (2 / 4) & \ding{51} & \ding{51} & \ding{55} \\
    find-in-sorted             & \ding{51} (3 / 4) & \ding{55} & \ding{55} & \ding{55} \\
    \rowcolor{lightgray}
    flatten                    & \ding{51} (4 / 4) & \ding{51} & \ding{51} & \ding{55} \\
    gcd                        & \ding{55} (0 / 4) & \ding{51} & \ding{55} & \ding{55} \\
    \rowcolor{lightgray}
    get-factors                & \ding{51} (1 / 4) & \ding{51} & \ding{51} & \ding{55} \\
    hanoi                      & \ding{55} (0 / 4) & \ding{51} & \ding{51} & \ding{55} \\
    \rowcolor{lightgray}
    is-valid-parenthesization  & \ding{51} (2 / 4) & \ding{51} & \ding{55} & \ding{55} \\
    kheapsort                  & \ding{55} (0 / 4) & \ding{51} & \ding{55}& \ding{55} \\
    \rowcolor{lightgray}
    knapsack                   & \ding{51} (1 / 4) & \ding{51} & \ding{51} & \ding{51} \\
    kth                        & \ding{55} (0 / 4) & \ding{55} & \ding{55} & \ding{55} \\
    \rowcolor{lightgray}
    lcs-length                 & \ding{55} (0 / 4) & \ding{55} & \ding{51} & \ding{55} \\
    levenshtein                & \ding{55} (0 / 4) & \ding{55} & \ding{55} & \ding{51} \\
    \rowcolor{lightgray}
    lis                        & \ding{55} (0 / 4) & \ding{55} & \ding{55} & \ding{51} \\
    longest-common-subsequence & \ding{55} (0 / 4) & \ding{51} & \ding{55} & \ding{55} \\
    \rowcolor{lightgray}
    max-sublist-sum            & \ding{55} (0 / 4) & \ding{51} & \ding{55} & \ding{55} \\
    mergesort                  & \ding{51} (1 / 4) & \ding{55} & \ding{55} & \ding{51} \\
    \rowcolor{lightgray}
    minimum-spanning-tree      & \ding{55} (0 / 4) & \ding{55} & \ding{51} & \ding{55} \\
    next-palindrome            & \ding{51} (1 / 4) & \ding{55} & \ding{51} & \ding{55} \\
    \rowcolor{lightgray}
    next-permutation           & \ding{55} (0 / 4) & \ding{55} & \ding{51} & \ding{55} \\
    pascal                     & \ding{51} (1 / 4) & \ding{55} & \ding{51} & \ding{55} \\
    \rowcolor{lightgray}
    possible-change            & \ding{51} (1 / 4) & \ding{51} & \ding{55} & \ding{55} \\
    powerset                   & \ding{55} (0 / 4) & \ding{51} & \ding{55} & \ding{55} \\
    \rowcolor{lightgray}
    quicksort                  & \ding{51} (1 / 4) & \ding{51} & \ding{55} & \ding{51} \\
    reverse-linked-list        & \ding{51} (2 / 4) & \ding{51} & \ding{55} & \ding{55} \\
    \rowcolor{lightgray}
    rpn-eval                   & \ding{55} (0 / 4) & \ding{55} & \ding{51} & \ding{51} \\
    shortest-path-length       & \ding{51} (1 / 4) & \ding{55} & \ding{55} & \ding{55} \\
    \rowcolor{lightgray}
    shortest-path-lengths      & \ding{55} (0 / 4) & \ding{55} & \ding{51} & \ding{55} \\
    shortest-paths             & \ding{51} (1 / 4) & \ding{55} & \ding{55} & \ding{55} \\
    \rowcolor{lightgray}
    shunting-yard              & \ding{51} (2 / 4) & \ding{55} & \ding{55} & \ding{55} \\
    sieve                      & \ding{55} (0 / 4) & \ding{51} & \ding{51} & \ding{55} \\
    \rowcolor{lightgray}
    sqrt                       & \ding{51} (1 / 4) & \ding{51} & \ding{51} & \ding{55} \\
    subsequences               & \ding{51} (1 / 4) & \ding{55} & \ding{51} & \ding{55} \\
    \rowcolor{lightgray}
    to-base                    & \ding{55} (0 / 4) & \ding{51} & \ding{55} & \ding{55} \\
    topological-ordering       & \ding{55} (0 / 4) & \ding{55} & \ding{51} & \ding{55} \\
    \rowcolor{lightgray}
    wrap                       & \ding{55} (0 / 4) & \ding{51} & \ding{55} & \ding{55} \\
    \hline
    \pmb{$\Sigma$}\textbf{ (Solved)}   & \textbf{19}   & \textbf{21}  & \textbf{19} & \textbf{7} \\
  \hline 
\end{tabular}
\end{table*}

\section{Results and Discussion}\label{sec:results}

In this section, we present the results of the comparison of ChatGPT, Codex, CoCoNut, and the standard APR approaches. 
We classify ChatGPT's answers and report on short discussions with the model.
Furthermore, we describe what we noticed while working with ChatGPT. 

\subsection{Automatic Bug Fixing Performance}\label{sec:performance}

Table~\ref{tab:results} shows the achieved results of ChatGPT, Codex, CoCoNut, and the dedicated APR approaches on the benchmark problems from QuixBugs. 
For the ChatGPT results, a checkmark (\ding{51}) indicates that a correct answer was given in at least one of the four runs for a benchmark problem. A cross (\ding{55}) indicates that no correct answer was given in any of the runs. In parentheses we additionally report the number of runs that led to a successful solution. For the results from the literature, a checkmark indicates that a correct bug fix is reported. A cross means that no successful bug fix is reported. 

We see that the results achieved by ChatGPT are similar to Codex in performance and outperform the standard APR approaches. Overall, we find bug fixes for $19$ benchmark problems with ChatGPT, $21$ are reported for Codex, $19$ for CoCoNut, and only $7$ for the standard approaches.  

The large gap in performance between the language model based approaches and the standard APR approaches can be explained by the fact that the latter usually just use a small test suite to define the problem, which can be easily overfitted. The authors of \cite{ye2021comprehensive} also report this problem. If only the test suite is considered for evaluation, the standard approaches would solve a total of $16$ benchmark problems. However, as in real-world applications only programs that work also on unseen inputs are usable, we have only adopted the $7$ generalizing problems from \cite{ye2021comprehensive} as correct.

If we take a closer look at the results for ChatGPT, we see that benchmark problems are often only solved in one or two runs. Only for the problems {\sc bucketsort} and {\sc flatten} ChatGPT finds a bug fix in all four runs. So ChatGPT seems to have a relatively high variance when fixing bugs. For an end-user, however, this means that it can be helpful to execute requests multiple times. 

Furthermore, it is not surprising that ChatGPT solves about the same number of problems as Codex, as ChatGPT and Codex are from the same family of language models.\footnote{\url{https://beta.openai.com/docs/model-index-for-researchers} (accessed January 18, 2023).} However, we still see potential for improvement for ChatGPT, as the given responses are often close to the correct solution (for a detailed classification of ChatGPT's responses see Section~\ref{sec:classification}). 

 Nevertheless, we are very strict in our evaluation and consider only patches as correct if the bug introduced by QuixBugs is actually identified and corrected. E.g., for some problems, ChatGPT suggests a complete re-implementation which is then bug-free. However, these are probably no real bug fixes, since the introduced bug is not localized. We assume that ChatGPT simply reproduced what it has learned here. Furthermore, we do not count a bug as fixed if additional changes suggested by ChatGPT introduce new errors that prevent the program from running properly. Moreover, by sending just a single request in this evaluation, we are not using the full potential of the dialogue system. Consequently, we take a closer look at how ChatGPT behaves when we interact more with the system and give it more information about the bug in Section~\ref{sec:dialog}.

\begin{figure*}[ht!]
  \centering
  \includegraphics[width=1.0\textwidth]{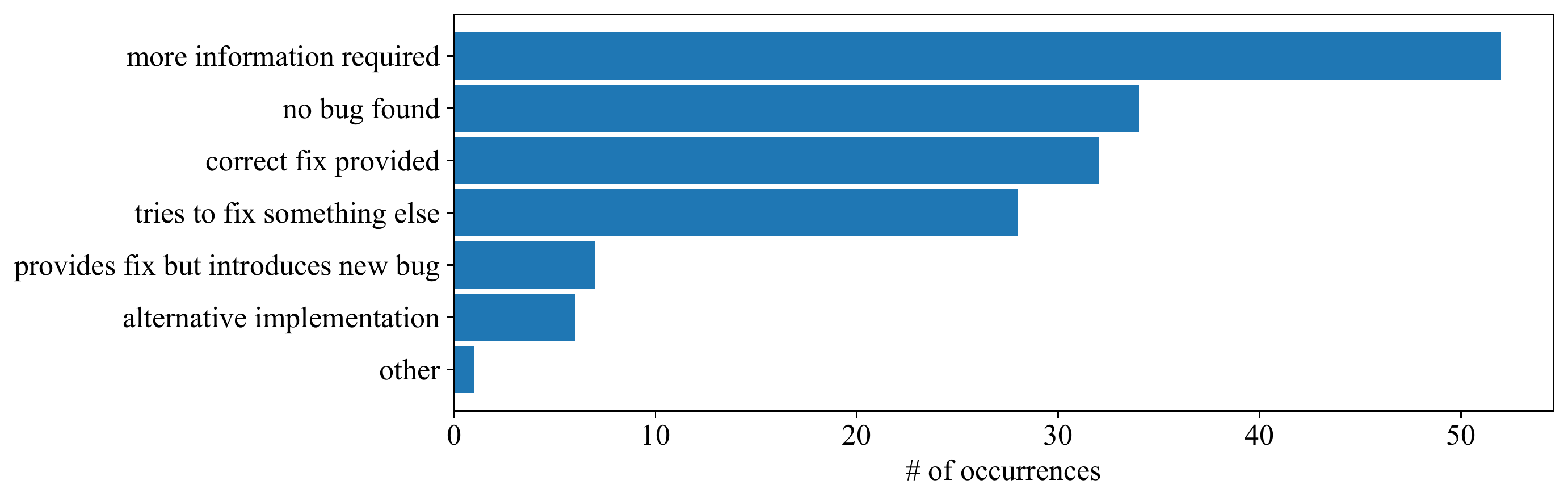}
  \caption{Number of occurrences of identified classes of ChatGPT answers given for the problems from QuixBugs.}
  \label{fig:barplot}
\end{figure*}

\begin{figure*}[!ht]
\lstset{
  basicstyle=\small,
  moredelim=[is][\textbf]{_}{_}
}
\begin{lstlisting}[frame=lines,numbers=left,xleftmargin=2.4em,framexleftmargin=2.3em]
_Request (1):_
Does this program have a bug? How to fix it?

def gcd(a, b):
    if b == 0:
        return a
    else:
        return gcd(a % b, b)

_Response (1):_
This is the Euclidean algorithm for finding the greatest common divisor (GCD) of two 
integers, a and b. It appears to be implemented correctly. If you are experiencing 
issues with this program, it is likely that the problem lies elsewhere in your code. 
Without more information about the bug you are experiencing, it is difficult to say 
how to fix it.

_Request (2):_
No, the function is not correct. E.g., for the input a = 4 and b = 5 I get the following 
error message: "RecursionError: maximum recursion depth exceeded in comparison".

_Response (2):_
The issue is that the function calls itself recursively, and in some cases, it never 
reaches the base case (b == 0), causing it to recurse indefinitely. The base case of 
the gcd algorithm is when the remainder is zero, but in this implementation, it's when 
b is zero. For example, when a=4 and b=5, gcd(4, 5) is called, which then calls 
gcd(4 % 5, 5) which is gcd(4, 5). This causes an infinite loop.

You can fix this by swapping the parameters in the recursive call.

def gcd(a, b):
    if a == 0:
        return b
    else:
        return gcd(b % a, a)

This way, if a is zero, the function returns b (which is the gcd of a and b), otherwise 
it calls itself with the new values of b % a and a.
\end{lstlisting}
\caption{A discussion with ChatGPT about the {\sc gcd} problem from QuixBugs. We inserted request and response markers in \textbf{bold} font to make the chat history more readable.}
\label{fig:example_gcd}
\end{figure*}

\begin{figure*}[!ht]
\lstset{
  basicstyle=\small,
  moredelim=[is][\textbf]{@}{@}
}
\begin{lstlisting}[frame=lines,numbers=left,xleftmargin=2.4em,framexleftmargin=2.3em]
@Request (1):@
Does this program have a bug? How to fix it?

def bitcount(n):
    count = 0
    while n:
        n ^= n - 1
        count += 1
    return count

@Response (1):@
I'm unable to tell if the program has a bug without more information on the expected 
behavior and the input that is causing the issue. [...]

@Request (2):@
The program should count the number of 1-bits in the binary encoding of n.

@Response (2):@
The function bitcount(n) appears to be counting the number of 1-bits [...]

To confirm this, you can test the function with some test cases to check if 
it produces the expected output:

test_cases = [0, 1, 2, 3, 4, 5, 6, 7, 8, 9, 10, 11, 12, 13, 14, 15]
expected_output = [0, 1, 1, 2, 1, 2, 2, 3, 1, 2, 2, 3, 2, 3, 3, 4]

@Request (3):@
The function does not work. For input 1 it seems to run into an infinite loop.

@Response (3):@
You are correct that the function will run into an infinite loop when the input is 1. 
[...]

def bitcount(n):
    if n == 0:
        return 0
    count = 0
    while n:
        n ^= n - 1
        count += 1
    return count

def bitcount(n):
    count = 0
    while n > 0:
        n ^= n - 1
        count += 1
    return count

def bitcount(n):
    return bin(n).count('1')

def bitcount(n):
    count = 0
    while n:
        if n & 1:
            count += 1
        n >>= 1
    return count
\end{lstlisting}
\caption{A discussion with ChatGPT about the {\sc bitcount} problem from QuixBugs. We inserted request and response markers in \textbf{bold} font to make the chat history more readable. In addition, we shortened ChatGPT's responses in some places to keep the example as small as possible. When we left out large text passages, we marked this in the text.}
\label{fig:example_bitcount}
\end{figure*}

\subsection{A Classification of ChatGPT's Answers}\label{sec:classification}

While working with ChatGPT, we noticed different types of responses that ChatGPT gave to our requests, especially when a bug could not be found. Therefore, we identified the different types of answers from ChatGPT for the benchmark problems from QuixBugs and analyzed their frequency. We identified the following classes of ChatGPT answers: 

\begin{itemize}
    \item \textbf{More information required}: Asks for more information on the program behavior to identify the bug.
    \item \textbf{No bug found}: Does not find a bug and states the program is working correctly.
    \item \textbf{Correct fix provided}: Provides the correct fix for the correct bug.
    \item \textbf{Tries to fix something else}: Does not find the intended bug and tries to fix or advise on something else that is not really a bug or adjusts for edge cases.
    \item \textbf{Provides fix but introduces new bug}: Provides the correct fix for the target bug but introduces a new bug somewhere else.
    \item \textbf{Alternative implementation}: Does not fix the bug but gives a working alternative implementation.
\end{itemize}

Figure~\ref{fig:barplot} shows the number of occurrences of identified classes of ChatGPT answers given for the problems from QuixBugs. 

We see that for most of our requests, ChatGPT asks for more information about the problem and the bug. With the second most number of answers given, we observe ChatGPT claiming that the given code snippet does not seem to have a bug. In both cases it might be useful to fully utilize the possibilities of the dialogue system ChatGPT offers, as further information might lead to a correct bug fix. 

Less often than the request for more information, we observe that ChatGPT fixes the bug but at the same time introduces new errors, or we see that ChatGPT not really addresses the bug correctly but suggests a completely new working re-implementation for the problem.

\subsection{A Discussion with ChatGPT}\label{sec:dialog}

In order to be able to compare ChatGPT with other systems in a standardized form, we have so far studied how ChatGPT behaves when making exactly one request (see Section~\ref{sec:performance}). However, a potentially powerful advantage of ChatGPT is that we can interact with the system in a dialogue to specify a request in more detail. This is useful, as both, a textual description as well as a definition given by a test suite, are possibly ambiguous and could be misleading. 

Consequently, we investigate for two benchmark problems how ChatGPT behaves in a conversation and if it is possible to find a working solution by discussing the problem with the system. We choose {\sc gcd} and {\sc bitcount} as benchmark problems because in our previous experiments the contained bug could not be fixed correctly for both problems. Furthermore, the problems consist of a relatively small number of code lines which allows us to discuss these problems in detail.

Figure~\ref{fig:example_gcd} shows an example discussion with ChatGPT about the {\sc gcd} problem (lines $1$--$8$). In the first response (lines $10$--$15$), ChatGPT does not present any solution. It asks for more information about the bug (we observed this behavior for many other problems, see Section~\ref{sec:classification}). Since the given function causes recursion issues for many possible inputs, we give ChatGPT an exact input example and the resulting error message from Python (lines $17$--$19$). By mentioning the recursion issue, the final response goes in the right direction and we get a correctly working patched version (lines $30$--$34$). 

In Figure~\ref{fig:example_bitcount} we see an example discussion with ChatGPT about the {\sc bitcount} problem (lines $1$--$9$). Again, ChatGPT asks for more information about the problem and for an input that causes an error (lines $11$--$13$). As follow-up request, we give ChatGPT a description of what the function should do (based on a code comment from QuixBugs) and ignore the request for an example input to see how ChatGPT reacts (lines $15$ and $16$). We can see in the following answer (lines $18$--$25$) that there is clearly a relation between ChatGPT's first and second answer because now we get an explanation of how we can test the function with some test inputs. We respond with a problem description for a test input and describe that there is probably an issue with an infinite loop (lines $27$ and $28$). ChatGPT responds with four code snippets where the first two (lines $34$--$48$) do not solve the problem with the infinite loop and the last two (lines $50$--$59$) are complete but working re-implementations which, however, not directly address the contained bug. It seems that ChatGPT simply returns functions here that somehow fit the content of the problem discussion, even though the test cases mentioned by ChatGPT show that the first two functions cannot work correctly. Also the bug is not simply fixed by replacing \texttt{\lstinline{n ^= n - 1}} with \texttt{\lstinline{n &= n - 1}} in the given function, but ChatGPT, as already mentioned, returns two complete re-implementations. However, both observations are not particularly surprising for a language model based approach. Nevertheless, the given answers would be useful for a programmer as they help to solve the problem.

\subsection{Systematic Follow-up Requests for ChatGPT}

Next, we conducted a study where we systematically discuss with ChatGPT. For those programs for which the contained bug was not correctly addressed by ChatGPT (see Table~\ref{tab:results}), we provide ChatGPT with a follow-up request giving a hint, as specified in Section~\ref{subsec:dialogue}. We report our results in Table~\ref{tab:results_remaining}. We use the same notation as before with the addition that a checkmark with an asterisk (\ding{51}*) defines that a solution was found without a follow-up request being necessary in this run. 

For $9$ benchmark problems we see that a more detailed description of the bug is helpful for ChatGPT. For $3$ benchmark problems no follow-up request was necessary in this run, since the bug was correctly addressed in the response given on our first request. 
Overall, adding a hint to ChatGPT vastly improves its performance, with $31$ out of $40$ problems solved. 
ChatGPT thus offers an exciting new way of approaching automated program repair.

\definecolor{lightgray}{gray}{0.9}
\begin{table}
  \centering
  \caption{Results achieved by ChatGPT with additional information given in a follow-up request for the unsolved benchmark problems (see Table~\ref{tab:results}).}
  \renewcommand{\arraystretch}{1.64}
  \label{tab:results_remaining}
  \begin{tabular}{lc}
    \hline
    \textbf{Benchmark problem}\;\;\;\;\;\;\;\;\;\;\; & \;\;\;\;\;\;\;\;\textbf{ChatGPT}\;\;\;\;\;\;\;\; \\
    \hline
    bitcount                   & \ding{51} \\
    \rowcolor{lightgray}
    depth-first-search         & \;\;\ding{51}* \\
    detect-cycle               & \;\;\ding{51}* \\
    \rowcolor{lightgray}
    gcd                        & \ding{51} \\
    hanoi                      & \ding{51} \\
    \rowcolor{lightgray}
    kheapsort                  & \ding{55} \\
    kth                        & \ding{51}  \\
    \rowcolor{lightgray}
    lcs-length                 & \ding{55} \\
    levenshtein                & \ding{51} \\
    \rowcolor{lightgray}
    lis                        & \ding{55} \\
    longest-common-subsequence & \ding{55} \\
    \rowcolor{lightgray}
    max-sublist-sum            & \ding{51} \\
    minimum-spanning-tree      & \ding{51} \\
    \rowcolor{lightgray}
    next-permutation           & \ding{51} \\
    powerset                   & \ding{51} \\
    \rowcolor{lightgray}
    rpn-eval                   & \ding{55} \\
    shortest-path-lengths      & \ding{55} \\
    \rowcolor{lightgray}
    sieve                      & \;\;\ding{51}* \\
    to-base                    & \ding{55} \\
    \rowcolor{lightgray}
    topological-ordering       & \ding{55} \\
    wrap                       & \ding{55} \\
    \hline
    \pmb{$\Sigma$}\textbf{ (Solved)}   & \textbf{9 (12)} \\
  \hline 
\end{tabular}
\end{table}

\section{Threats to Validity}\label{sec:threats} 

It is worth noting that ChatGPT is currently under active development. 
During our study there was a major update to it, which might have influenced our results.
Although we observed repairability rates before and after the update to be similar. However, future releases might yield different results.
Furthermore, ChatGPT allows for conversation with its users.
Asking a different question than the ones presented in this study could potentially have a different impact on results.
To mitigate this threat to validity, we conducted a pre-study, varying the questions asked. 
We noted no significant influence on the results.
Moreover, the results might vary depending on the programming language, size of the benchmarks, and the number of queries issued.
To mitigate these threats, we chose a standard benchmark set and targeted Python -- the most popular programming language.\footnote{\url{https://www.tiobe.com/tiobe-index/} (accessed January 18, 2023).} 
The classification of the results was done manually and therefore represents the subjective assessment of the authors. To enable a verification of our results, we made our conversations with ChatGPT available online.

\section{Conclusions and Future Work}\label{sec:conclusions}
To support programmers in finding and fixing software bugs, several automated program repair (APR) methods have been proposed. ChatGPT, a recently presented deep learning (DL) based dialogue system, can also make suggestions for improving erroneous source code. 
However, so far the quality of these suggestions has been unclear. Therefore, we compared in this work the automatic bug fixing performance of ChatGPT with that of Codex and several dedicated APR approaches.

We find that ChatGPT has similar performance to Codex and dedicated DL-based APR on a standard benchmark set. 
It vastly outperforms standard APR methods (19 vs. 7 out of 40 bugs fixed).
Using ChatGPT's dialogue option and giving the system more information about the bug in a follow-up request boosts the performance even further, giving an overall success rate of $77.5$\%. 
This shows that human input can be of much help to an automated APR system, with ChatGPT providing means to do so.

Despite its great performance, the question arises whether the mental cost required to verify ChatGPT answers outweighs the advantages that ChatGPT brings.
Perhaps incorporation of automated approaches to provide ChatGPT with hints as well as automated verification of its responses, e.g., through automated testing,  would yield ChatGPT to be a viable tool that would help software developers in their daily tasks.

We hope our results and observations will be helpful for future work with ChatGPT.

\section*{Acknowledgments}

This work was partially supported by UKRI EPSRC grant no. EP/P023991/1.

\bibliographystyle{IEEEtran}
\bibliography{bibliography}

\end{document}